\documentstyle[prl, aps, amssymb
]{revtex}
\begin{document}
\draft
\wideabs{
\title{Spin-Dynamics in the Carrier-Doped $S=\frac{1}{2}$ Triangular-Lattice 
of Na$_{x}$CoO$_{2}$-yH$_{2}$O
}
\author{F.L. Ning $^{a}$, T. Imai$^{a,b}$, B.W. Statt $^{c}$, and F.C. 
Chou$^{d}$}
\address{$^{a}$Department of Physics and Astronomy,
McMaster University, Hamilton, Ontario L8S 4M1, CANADA}
\address{$^{b}$Canadian Institute for Advanced Research, Toronto, Ontario 
M5G 1Z8, CANADA}
\address{$^{c}$Department of Physics,
University of Toronto, Toronto, Ontario M5S 1A7, CANADA}
\address{$^{d}$Center for Materials Science and 
Engineering, M.I.T., Cambridge, MA 02139}
\date{\today}
\maketitle
\begin{abstract}
We probed the local electronic properties of the mixed-valent
Co$^{+4-x}$  triangular-lattice  in 
Na$_{x}$CoO$_{2}$-yH$_{2}$O by $^{59}$Co NMR.  
We observed two distinct types of Co sites for $x\geq \frac{1}{2}$, but the 
valence seems averaged out for $x\sim\frac{1}{3}$.  Local 
spin fluctuations exhibit qualitatively the same
trend down to $\sim$100 K regardless 
of the carrier-concentration $x$, and hence the nature 
of the electronic ground state.  A canonical Fermi-liquid 
behavior emerges below $\sim$100 K only for $x\sim\frac{1}{3}$.
\end{abstract}
}
There is wide-spread speculation that the water-intercalated
Na$_{0.3}$CoO$_{2}$-1.3H$_{2}$O ($T_{c}\sim4.5K$)\cite{Takada} represents the first 
example of the long-sought RVB (Resonating Valence Bond 
\cite{Anderson}) superconductor 
in a carrier-doped $S=\frac{1}{2}$ 
{\it triangular-lattice}\cite{Baskaran}.  The symmetry of the order-parameter appears 
to be unconventional\cite{Fujimoto}.  The parent 
compound of the superconducting Na$_{0.3}$CoO$_{2}$-1.3H$_{2}$O is a mixed-valent 
Na$_{x}$CoO$_{2}$ with sheets of 
Co$^{4+}$ ions ($S=\frac{1}{2}$) 
arranged in a triangular-lattice.  Na$^{+}$ ions dope additional electrons 
with probability $x$ \cite{Terasaki}.  The electronic and magnetic 
phase diagram of Na$_{x}$CoO$_{2}$ is quite rich \cite{Terasaki,Foo,Sugiyama,Uemura,Keimer} 
because of the interplay between doped-carriers and inherently 
frustrated spins on triangular-lattice (see Fig.1), and includes 
commensurate antiferromagnetism ($x=0.82$\cite{Uemura} and $0.5$\cite{Keimer}), 
SDW ($x\sim 0.75$\cite{Sugiyama}), 
and Fermi-liquid state ($x\sim 0.31$\cite{Foo}).  
Among the key issues 
are:  does charge order(s) exist in carrier-doped Co triangular-lattice, as 
hinted by earlier magnetic resonance\cite{Uemura,Keimer,Ray} and bulk 
studies\cite{Foo}?
How do Co spin-fluctuations evolve with $x$? 
Is the ground state of $\frac{1}{3}$ doped triangular-lattice a Fermi-liquid 
or ferromagnetic\cite{Baskaran}?  Do exotic phases ({\it e.g.} pseudo-gap\cite{Timusk} and 
stripe \cite{Tranquada}) exist,
in analogy with high $T_{c}$ cuprates?  

In this {\it Letter}, we will shed new light on the anomalous behavior 
of the carrier-doped Co triangular-lattice 
in Na$_{x}$CoO$_{2}$-yH$_{2}$O based on systematic  
$^{59}$Co NMR measurements. 
The main thrust of our approach is that we can  {\it separately} observe and characterize 
Co ions in different valence states.  From the NMR 
lineshape (Fig.2), local uniform spin 
susceptibility (as measured by $^{59}$Co NMR Knight shift $^{59}K$, Fig.3), and 
local spin-fluctuations (as measured by nuclear spin-lattice 
relaxation rate $\frac{1}{T_{1}}$, Fig.4), we establish that 
two types of mutually coupled Co ions 
with different local electronic properties exist in $x\geq$0.5.  
Furthermore, we show that low-frequency spin-fluctuations 
exhibit qualitatively the same trend down to $\sim$100 K for 
all Na concentrations $x$ despite their different 
electronic ground states.  A canonical
Fermi-liquid behavior emerges below 100 K only 
for $x$=0.3, and water-intercalation enhances spin-fluctuations near 
$T_{c}$. 

$^{59}$Co nuclei have spin $I$=$\frac{7}{2}$ (nuclear gyromagnetic 
ratio $\gamma_{n}$=10.053 MHz/Tesla), and seven NMR transitions from 
the $I_{z}$=$m$ to $m+1$ state are expected for each different type of 
Co site\cite{Slichter}. We show representative lineshapes observed in the external magnetic 
field $H_{ext}$=8 Tesla in Fig.2.  All samples are single-crystals 
\cite{Chou} 
except for $x=0.72$; the latter is a ceramic sample synthesized by the standard 
{\it fast-heating} 
method \cite{Motohashi}, and magnetically aligned in stycast 1266.  The lineshapes for 
$x=0.75$, 
$0.72$, $0.67$, and 
$0.5$ clearly establish the presence of at least two different types of 
$^{59}$Co NMR lines.  They have different central resonance frequencies,
 {\it e.g.} $\nu_{-\frac{1}{2},+\frac{1}{2}}\sim$82 MHz 
at A'-sites and $\sim$86 MHz 
at B'-sites in Na$_{0.72}$CoO$_{2}$.  In addition, the splitting, $\nu_{a}$, between 
the central and satellite transitions due to nuclear quadrupole interaction 
with the local Electric-Field-Gradient (EFG)\cite{Slichter} is 
different, {\it e.g.} $\nu_{a}\sim$560 kHz at A'-sites and $\nu_{a}\sim$650 kHz 
at B'-sites.  

In the absence of hyperfine interactions with Co electrons, the 
$^{59}$Co NMR frequency would be $\gamma_{n}H_{ext}$=80.4 MHz in $H_{ext}$=8 Tesla.  
It is primarily the hyperfine magnetic 
field at each Co site that shifts the central resonance frequency $\nu_{-\frac{1}{2},+\frac{1}{2}}$, 
and the former 
has a linear relation to the 
local magnetic susceptibility $\chi_{spin}^{i}$ at the {\it i}-th Co site.  
Generally, one can gain information on $\chi_{spin}^{i}$ through the NMR Knight shift 
$^{59}K=(\nu_{-\frac{1}{2},+\frac{1}{2}}-\gamma_{n}H_{ext})/\gamma_{n}H_{ext}$
\cite{Slichter} as presented in Fig.3\cite{CE}, where 
\begin{equation}
          ^{59}K= ^{59}K_{spin}+^{59}K_{orb}=\sum_{i=0\sim 
          6}H_{i}\chi_{spin}^{i}+^{59}K_{orb}.
\label{K2}
\end{equation}
$^{59}K_{spin}$ is the spin contribution to the NMR Knight shift, 
$H_{0}$ is the hyperfine interaction between the observed Co nuclear-spin and the 
Co electron-spin at the same site, and $H_{1\sim6}$ represents the hyperfine interaction with electron 
spins at six nearest-neighbor Co sites.  The orbital contribution is 
usually temperature-independent, and is
estimated to be $^{59}K_{orb}=(2\sim3)$\% below 250K based on preliminary $K$ 
vs. $\chi$ analysis\cite{Ning}.
 
 In Fig.4, we summarize $\frac{1}{T_{1}T}$, the $^{59}$Co 
 nuclear spin-lattice relaxation rate $\frac{1}{T_{1}}$ divided by 
 temperature $T$.  
 $\frac{1}{T_{1}T}$ measures the low-frequency spectral weight of the local Co spin-fluctuations 
 at the experimental frequency $\nu_{n} = 78\sim86$ MHz (depending on 
 the particular transition we employ).  
 Theoretically, the spin contributions to $\frac{1}{T_{1}T}$ may be 
 written using the imaginary part of the Co dynamical electron 
spin-susceptibility  ${\chi''({\bf q},\nu_{n})}$ as\cite{Moriya},
 \begin{equation}
          \frac{1}{T_{1}T} = 
          \frac{2\gamma_{n}^{2}k_{B}}{g^{2}\mu_{B}^{2}}\sum_{{\bf q}} | A({\bf q}) |^{2} \frac{\chi''({\bf 
	q},\nu_{n})}{\nu_{n}},
\label{T1}
\end{equation}
where $A({\bf q})=H_{0}+\sum_{i=1\sim 6}H_{i} e^{i{\bf q}{\bf r}_{i}}$ is the 
wave-vector {\bf q} dependent hyperfine {\it form 
factor}\cite{Moriya,Shastry}.

It is very important to realize that the spin contribution to the
Knight shift show identical temperature dependence at two different Co sites 
in Na$_{0.75}$CoO$_{2}$ (A- and B-sites), as well as in Na$_{0.72}$CoO$_{2}$ (A' 
and B'-sites), despite the factor 6 to 8 difference in their 
magnitude.  To underscore this point, in Fig.3 we plotted $6.05*[^{59}K_{c}(A)-1.1\%]$
and $8.43*[^{59}K_{c}(A')-1.65\%]$ to match with $^{59}K_{a}$ at B- 
and B'-sites, respectively.  In Fig.4, we also demonstrate that $\frac{1}{T_{1}T}$ at B- and B'-sites 
show identical behavior with that at A- and A'-sites, respectively\cite{B}.
 These results are not consistent with a simple phase-separation 
picture, and imply that the strongly 
 magnetic B- and B'-sites are 
electronically coupled with the less magnetic 
A- and A'-sites, respectively.  Furthermore, the ratio of the integrated NMR intensity between the 
strongly magnetic (B- and B'-sites) and less magnetic Co sites (A- and 
A'-sites) is consistent with $1-x$ : $x$ after careful 
corrections for the fast transverse relaxation time $T_{2}$ at the B- 
and B'-site.  
These findings strongly suggest that 
a charge-order leads to two mutually coupled Co-sites:
 A- and A'-sites are less magnetic Co$^{3+}$-like ions with $S\sim0$, while 
B- and B'-sites are strongly magnetic Co$^{4+}$-like ions with 
$S\sim\frac{1}{2}$.

In passing, notice that the saturation below $\sim20$K of $^{59}K(A)$ and 
$^{59}K(B)$ (both in Fig.3) and SQUID data (not shown) in our Na$_{0.75}$CoO$_{2}$
 is consistent with the SQUID data reported by Foo et 
al.\cite{Foo} for their Na$_{0.75}$CoO$_{2}$ sample with 
accurately calibrated Na content.  Our NMR and SQUID data showed no evidence for 
a bulk magnetic phase transition in this Na$_{0.75}$CoO$_{2}$ 
crystal.  On the other hand, an earlier  $\mu$SR work 
reported \cite{Sugiyama} a SDW phase transition below $\sim20$K for 
$\sim$20\% volume fraction of a ceramic 
``Na$_{0.75}$CoO$_{2}$'' sample prepared by {\it fast heating} 
method\cite{Motohashi}.  Interestingly, our Na$_{0.72}$CoO$_{2}$ ceramic 
sample, 
prepared by the same {\it fast heating} method, showed a continuous 
increase of SQUID data through $\sim20$K 
following a Curie-Weiss behavior, and we 
found a variety of NMR 
signatures of a magnetic instability (e.g. a dramatic magnetic NMR line 
broadening of B'-line below $\sim$30K, 
accompanied by temperature-dependent line-splitting of A'-line and Na NMR 
line\cite{Ning}).  The 
Na-content of our 
ceramic sample was determined to be 0.72$\pm$0.02 based on the $c-$axis 
lattice-constant, which is a monotonous function of Na content 
\cite{Foo}.  The complete details will be reported elsewhere\cite{Ning}.

In the antiferromagnetic phase Na$_{0.5}$CoO$_{2}$ \cite{Uemura}, we also observed 
two types of Co NMR signals, E-sites and F-sites, 
with the integrated-intensity ratio 0.5:0.5.  
This is again consistent with a charge-order, but with a geometrical configuration different 
from Na$_{0.72-0.75}$CoO$_{2}$.  Based on the magnitude of 
$\frac{1}{T_{1}T}$, 
we may identify the 
E- and F-sites as the weakly and strongly magnetic Co sites, 
respectively.  However, the valence of E- and F-sites may not be very different 
from the nominal averaged value of 3.5, 
as the difference in the magnitude of $\frac{1}{T_{1}T}$ is at most 
$\sim$3 and less dramatic than the case of Na$_{0.72-0.75}$CoO$_{2}$.  
We also found that the $^{59}$Co NMR satellite lineshape 
begins to show additional line-splitting below 
88 K \cite{Ning} where 
SQUID data exhibit a kink\cite{Foo,Chou}.  This suggests that the configuration of 
charge-order changes below 88 K.  We note that a recent structural 
study at 
3.5K suggests a zig-zag chain structure with two distinct Co sites 
\cite{Huang}.  A large distribution in the magnitude of $\frac{1}{T_{1}T}$ 
prevented us from defining it properly below $\sim$88 K.

In metallic Na$_{0.67}$CoO$_{2}$, we observed only sharp 
D- and D$'$-lines above $\sim$30 K, and their NMR properties were very similar.  
With decreasing temperature, broad C-sites emerge while the linewidth of the D- and D$'$-sites broadens.  
$\frac{1}{T_{1}T}$ at D-sites is similar to that at the weakly magnetic 
A-sites in Na$_{0.75}$CoO$_{2}$.  
The ratio of the integrated-intensity between C- and D-, D$'$-sites is not consistent with 0.33:0.67 
nor 0.67:0.33.  The Co sites with magnetic moments
may be missing in the Na$_{0.67}$CoO$_{2}$ lineshape due to their fast relaxation times.

In contrast with the results for $x\geq 0.5$, we have detected 
essentially only one class 
of $^{59}$Co NMR lines, G and G$'$, 
in metallic Na$_{0.3}$CoO$_{2}$ at 4.2 K. 
$\nu_{a}$, $^{59}K$, and $\frac{1}{T_{1}T}$ were nearly identical at 
both sites.  Thus the valence of all 
Co ions in the ground state of $x\sim\frac{1}{3}$ doped Na$_{0.3}$CoO$_{2}$ seems
averaged out.  The special 
character of the electronic state at the $x\sim\frac{1}{3}$ doping may 
be induced by the geometry, as dynamic singlet formation would be 
energetically advantageous, see Fig.1(b).  However, we cannot rule out the possibility 
that changes in the nature and extent of Na$^{+}$ vacancy 
order may be driving the changes of electronic properties across $x=0.5$.  

Now we turn our attention to the low-frequency Co spin-dynamics.  
One of the most interesting findings of this study is that $\frac{1}{T_{1}T}$ shows qualitatively 
the same decrease with temperature down to $\sim$100 K at all types of Co sites in all concentrations, 
regardless of the nature of the electronic properties below $\sim$100 K 
and of the behavior of $^{59}K$ (i.e. the ${\bf q}$=${\bf 0}$ behavior of $\chi_{spin}^{i}$).  
For example: $\frac{1}{T_{1}T}$ at A-, A'-, and D-sites decreases 
$\sim$30\% from 250K to 
100K, where $\chi_{spin}^{i}$ shows an increase following a Curie-Weiss 
law; in $x=0.3$, $\frac{1}{T_{1}T}$ decreases $\sim$40\%  in 
the same temperature range while $\chi_{spin}^{i}$ also {\it decreases}.  
Even more interesting is the fact that $\frac{1}{T_{1}T}$ above 100K at G-sites in 
Na$_{0.3}$CoO$_{2}$ is roughly the average of $\frac{1}{T_{1}T}$ at weakly 
magnetic E- and strongly 
magnetic F-sites in Na$_{0.5}$CoO$_{2}$.  
Furthermore, our preliminary results for water-intercalated, superconducting 
single-crystal Na$_{0.3}$CoO$_{2}$-1.3H$_{2}$O show identical $\frac{1}{T_{1}T}$
to that of Na$_{0.3}$CoO$_{2}$ above $\sim$100 K.  The identical behavior indicates that 
the weaker inter-Co-layer coupling along the c-axis caused by 
water-intercalation does not affect the behavior of Co spin 
fluctuations above $\sim$100 K.  

Thus our $\frac{1}{T_{1}T}$ results establish qualitatively the same
 suppression in the carrier-doped 
$S=\frac{1}{2}$ triangular-lattice with a comparable temperature 
scale(s) of the 
order of $\sim$100 K or higher, regardless of the carrier concentration $x$ and the 
inter-layer coupling.  Interestingly, the intra-layer Co-Co exchange 
interaction is estimated to be 
$J=$140$\sim$280 K for Na$_{x}$CoO$_{2}$-yH$_{2}$O by Wang et 
al.\cite{Baskaran}.      
This reminds us that the Cu spin-lattice relaxation rate 
in the high-$T_{c}$ cuprate La$_{2-x}$Sr$_{x}$CuO$_{4}$ exhibits a 
universal value set entirely by the Cu-Cu exchange interaction 
$J\sim$1500 K above $\sim$500 K \cite{imai}.  Perhaps $J$ is the single 
dominant parameter of $\frac{1}{T_{1}T}$ in the present case, too.  

According to eq.(2), our $\frac{1}{T_{1}T}$ data imply that the ${\bf q}$-integral of the low-frequency 
components of Co spin-fluctuations, weighted with the form factor 
$|A({\bf q})|^{2}$, 
decreases with temperature down to $\sim$100 K.  Since the total 
magnetic moment must be 
conserved, a possible scenario is that some spectral weight is being shifted to higher energies, 
in analogy with the pseudo-gap behavior in under-doped high $T_{c}$ 
cuprates\cite{Timusk}.  
Alternatively, the growth of the short-rage order of Co spins below 
$T\sim J$ may result 
in the pile-up of a considerable fraction of the spectral weight of 
$\chi''({\bf q},\nu_{n})$ to a ${\bf q}$-region 
where the form factor $|A({\bf q})|^{2}$ happens to be small 
\cite{A(q)}.  

Surprisingly, despite qualitatively the same trend in spin dynamics for all 
concentrations $x$ down to $\sim$100 K, both $\frac{1}{T_{1}T}$ and 
$^{59}K$ level off below $\sim$100K only for $x$=0.3, i.e. Korringa behavior \cite{Slichter}.  
This evidences for 
the emergence of a low-temperature 
canonical Fermi-liquid behavior for $x\sim\frac{1}{3}$ below $\sim$100 
K, and is consistent with the  emergence of $\sim T^{2}$ behavior of resistivity 
observed only for $x\sim\frac{1}{3}$ below $\sim$30K\cite{Foo}.  
The large low-temperature specific-heat 
$\gamma\sim$10 mJ/K$^{2}$-mol implies 
a substantial mass-enhancement by a factor 4$\sim$7\cite{Cho}.  Quite 
remarkably, water-intercalation alters the 
temperature dependence of $\frac{1}{T_{1}T}$ for the $x\sim\frac{1}{3}$ 
doping below and only below $\sim$100 K.  
Somehow the 
reduced inter-layer coupling between Co sheets appears to introduce a new temperature 
(and/or energy) scale in the low temperature Fermi-liquid state in $x\sim\frac{1}{3}$.

To conclude, we demonstrated the presence of two-types of Co-sites in the carrier-doped Co triangular-lattice 
of Na$_{x}$CoO$_{2}$-yH$_{2}$O for $x\geq0.5$ by probing the Co layers
directly with $^{59}$Co NMR.  The temperature 
dependences of $^{59}K$ and $\frac{1}{T_{1}T}$ suggest 
these sites are electronically coupled to each other.  
We also showed semi-quantitatively the same trend of spin-fluctuations 
above $\sim$100 K for a variety carrier concentrations 
from $x=$0.3 to 0.75 with $y=0$, and for varying strength of inter-layer coupling 
for a fixed $x=0.3$ (i.e. $y=0$ and $y=1.3$).  Our $\frac{1}{T_{1}T}$ results 
suggest the presence of a certain energy scale that shows little dependence on 
$x$ and $y$.  We also demonstrated that a canonical Fermi-liquid behavior 
emerges at low temperatures only for $x\sim\frac{1}{3}$ but 
water-intercalation alters the low temperature spin dynamics.

We thank Y.S. Lee, J.H. Cho, P.A. Lee, T. Timusk, J. Hwang, and H. Alloul for 
discussions.  B.W.S. (deceased) was on leave at McMaster during this study.  
This work was supported by NSERC, NEDO, and CIAR at McMaster, and 
by NSF-02-13282 at MIT. 

% now the references. delete or change fake bibitem. delete next three
%   lines and directly read in your .bbl file if you use bibtex.

% \begin{figure}
% \caption{}
% \label{}
% \end{figure}

\begin{figure}
\caption{ (a) Schematics of the undoped ($x=0$) Co$^{4+}$ triangular-lattice 
with $S=\frac{1}{2}$.  
When the nearest-neighbor exchange interaction $J$ tends to align two $S=\frac{1}{2}$ spins 
(\#1 and \#2) in an anti-parallel configuration, frustration would result 
for spin \#3.  
(b) For finite values of $x$ ($x=\frac{1}{3}$ in this panel), extra electrons 
could help local-singlet formation, and alter 
spin-correlations dramatically.}
\label{triangles}
\end{figure}

\begin{figure}
\caption{$^{59}$Co NMR lineshape with $H_{ext}=8$ Tesla applied 
along a-axis.  (The lineshapes for Na$_{0.72}$CoO$_{2}$ and 
Na$_{0.3}$CoO$_{2}$-$1.3$H$_{2}$O were 
obtained with $H_{ext}$ applied within ab-plane for aligned ceramics 
and single crystals, respectively.  The lineshapes for $x$=0.75 and 
0.72 are corrected for $T_{2}$.)  The arrows indicate 
the central transition $\nu_{-\frac{1}{2},+\frac{1}{2}}$ at various Co 
sites. For $x$=0.75, 0.72, and 0.67, 
$^{23}$Na NMR line was superposed above $\sim$88 MHz.  
Some of the furthest satellite transitions are not shown.
}
\label{Spectra}
\end{figure}

\begin{figure}
\caption{The $^{59}$Co NMR Knight shifts 
$^{59}K_{a}$ at B, B', D, F, G, and H-sites.  (Note that $^{59}K_{orb}\sim 2-3\%$ is {\it not} 
subtracted.)  Also plotted are $^{59}K_{c}$ measured at A- and 
A'-sites scaled to match with B- and B'-sites (see the main text).}
\label{Kfig}
\end{figure}

\begin{figure}
\caption{The temperature dependence of $\frac{1}{T_{1}T}$.
}
\label{T1fig}
\end{figure}

% \begin{table}
% \caption{}
% \label{}
% \end{tabular}
% \end{table}


\begin{references}
\bibitem{CE}$^{59}K(C)$ and $^{59}K(E)$ (both $\sim$2\%) were 
difficult to measure due to their proximity to other lines.
\bibitem{A(q)}Notice, however, that the ${\bf q}$-dependence of $|A({\bf q})|^{2}$ is sensitive to the hyperfine 
couplings $H_{i}$, which in turn are sensitive to charge-order 
configurations for different $x$.
\end{references}
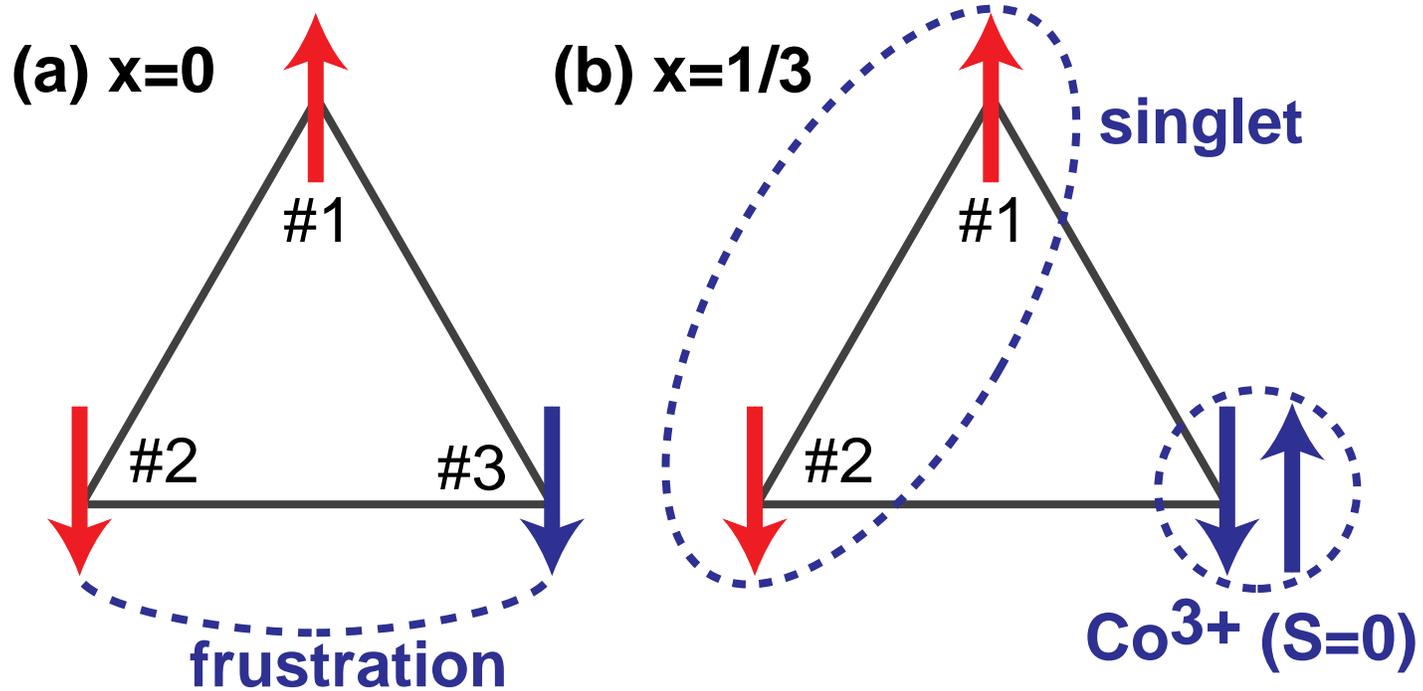
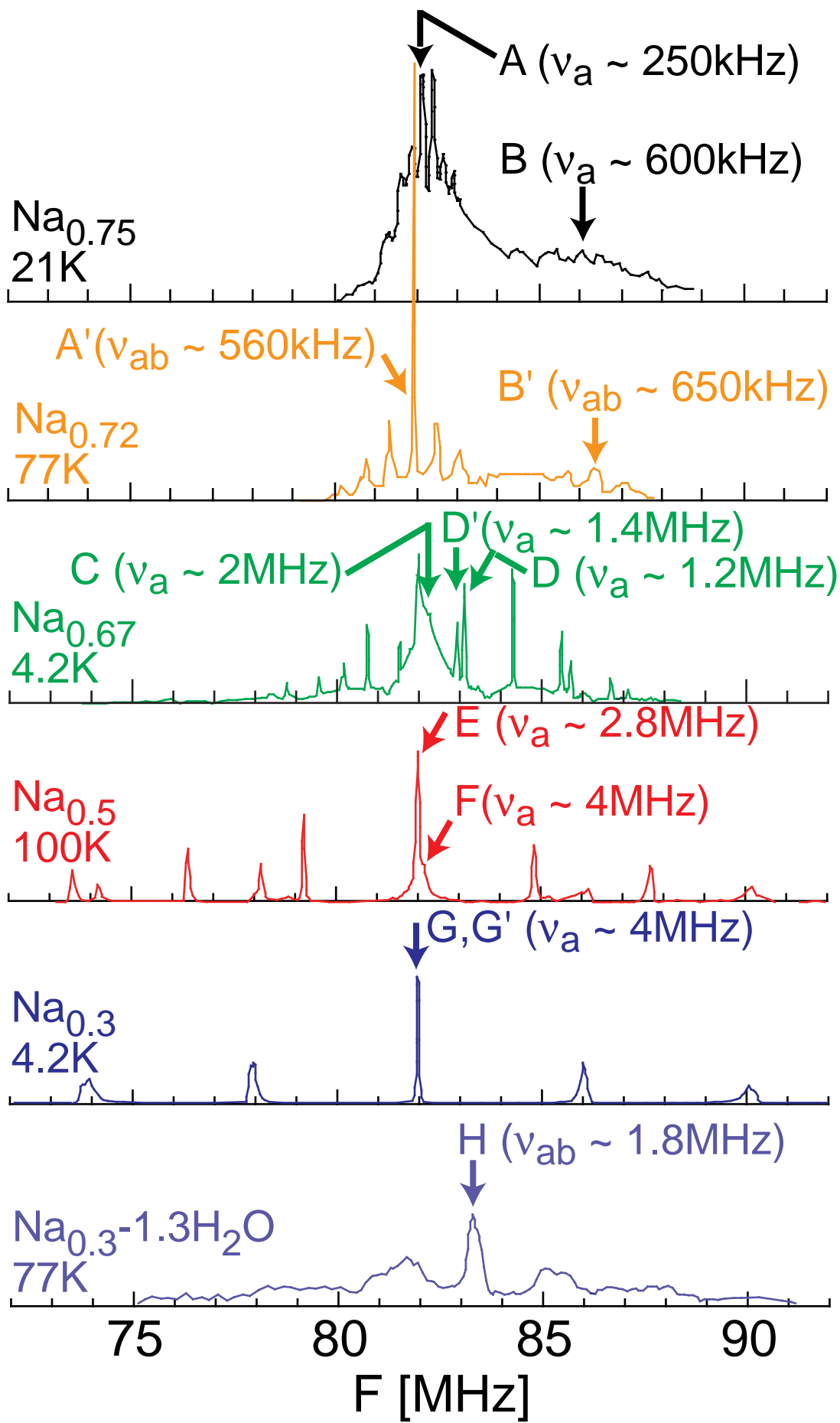
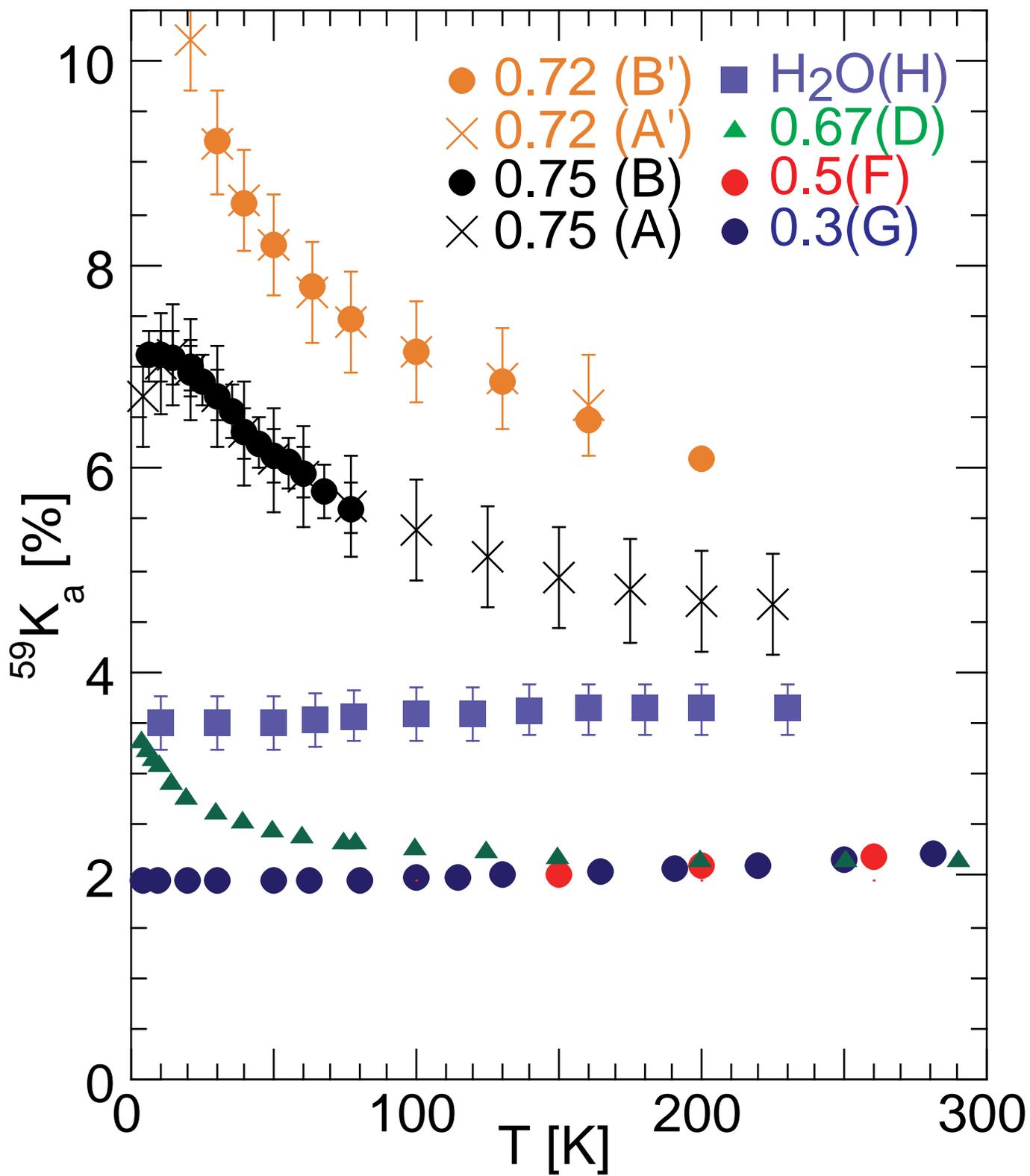
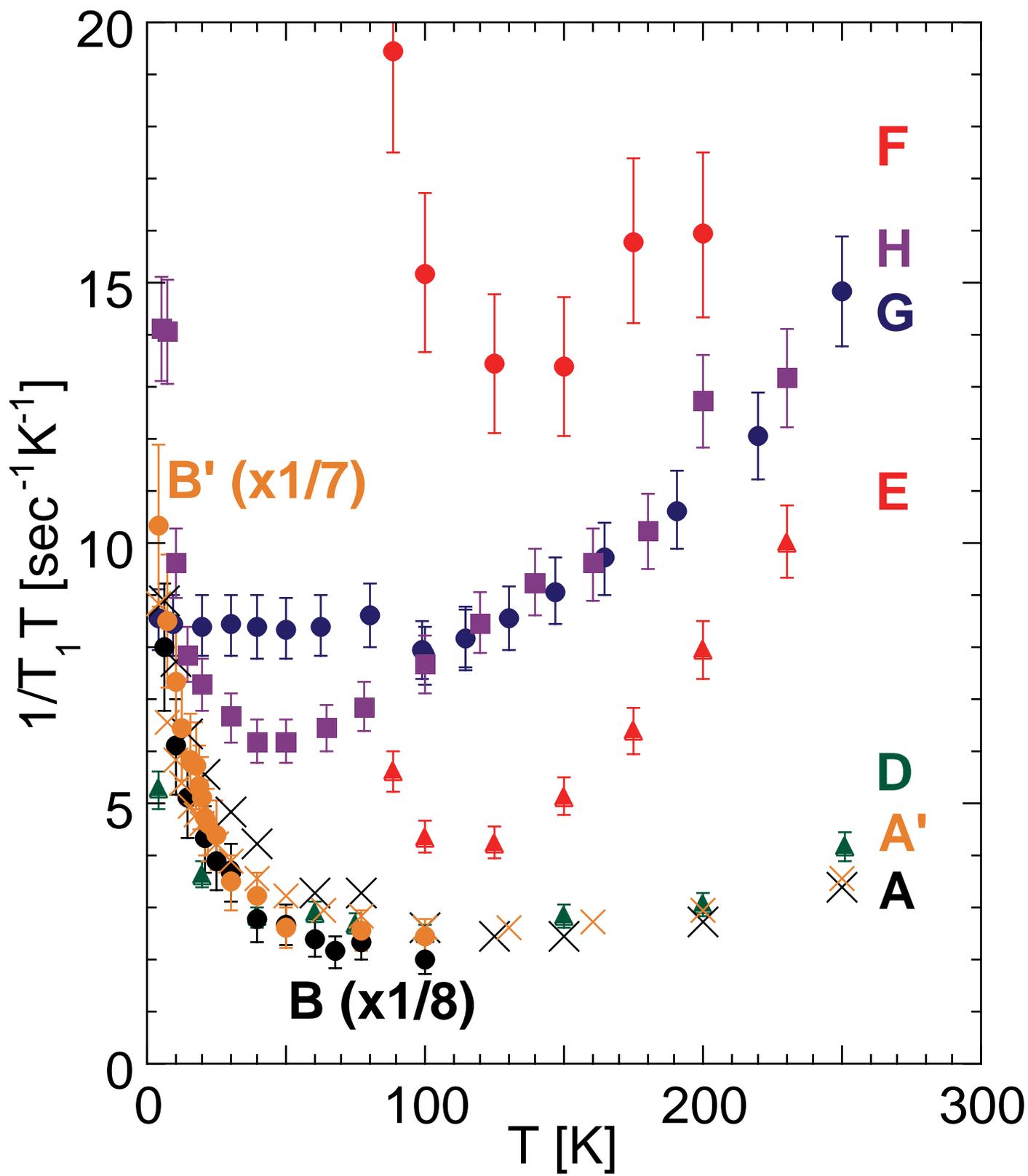
\end{document}